\documentclass[twocolumn,showpacs,preprintnumbers,amsmath,amssymb]{revtex4}
\usepackage{tabularx,graphicx}\begin{document}
\newcommand{\beq}{\begin{equation}}
\newcommand{\eeq}{\end{equation}}
\newcommand{\beqn}{\begin{eqnarray}}
\newcommand{\eeqn}{\end{eqnarray}}
\newcommand{\bmath}{\begin{subequations}}
\newcommand{\emath}{\end{subequations}}
\title{Electrodynamics of  spin currents in superconductors}
\author{J. E. Hirsch }
\address{Department of Physics, University of California, San Diego\\
La Jolla, CA 92093-0319}

\begin{abstract}

In recent work we formulated a new set of electrodynamic equations for superconductors as an alternative to the conventional London equations,  compatible with
the prediction of the theory of hole superconductivity that superconductors
expel negative charge from the interior towards the surface. Charge expulsion results  
in a macroscopically inhomogeneous charge distribution and an electric field in the interior, and because of this a spin current is expected to exist.
Furthermore, we have recently shown that a dynamical explanation of the Meissner effect in superconductors leads to the prediction that a spontaneous spin current
exists near the surface of superconductors (spin Meissner effect). 
In this paper we extend the  electrodynamic equations proposed earlier  for the charge density and charge current   to describe also  the  space and time dependence
of the spin density and spin current. This allows us to determine the magnitude of the expelled negative charge and interior electric field as well as of the spin current 
in terms of other measurable properties of superconductors. We also provide a 'geometric' interpretation of the difference between type I and type II superconductors,
discuss the relationship between our model and Slater's seminal work on superconductivity,
and discuss the magnitude of the expected novel effects for elemental and other superconductors.

\end{abstract}
\maketitle

\section{Introduction}
The theory of hole superconductivity predicts that superconductors expel negative charge
from the interior towards the surface\cite{pla01,sc0,prb03}, resulting in the existence of an electric field in the interior
of superconductors. To account for this fact we have proposed a new set of 
electrodynamic equations for superconductors\cite{prb03,edyn} in place of the conventional London equations\cite{london}.
The new equations predict  a charge density $\rho(\vec{r})$ in the interior
of  superconductors that satisfies the differential equation (in the static case)
\beq
\nabla^2\rho(\vec{r})=\frac{1}{\lambda_L^2}( \rho(\vec{r})-\rho_0)
\eeq
with $\rho_0$ a $positive$ constant. This equation  (together with a similar equation for the electric potential) predicts that the charge density in the deep interior of a superconducting body is
$\rho_0$, and that an excess of negative charge density  $\rho_-$ exists within a
London penetration depth ($\lambda_L$) of the surface. It also gives rise to an electric field in the interior
of superconductors that points towards the surface. For a cylindrical geometry and cylinder radius $R>>\lambda_L$,
charge conservation implies that
\bmath
\beq
\rho_0=-\frac{2\lambda_L}{R}\rho_-
\eeq
and for a sphere of radius $R>>\lambda_L$
\beq
\rho_0=-\frac{3\lambda_L}{R}\rho_-.
\eeq
\emath
In both cases, the electric field increases linearly away from the center, attains its maximum value 
\beq
E_m=-4\pi\lambda_L\rho_-
\eeq 
at a distance $\lambda_L$ from the surface, and drops to zero at the surface.

Energetic arguments\cite{prb03} indicate that 
$\rho_-$ and $E_m$ should be independent of the size of the sample and that $E_m$ should be related to the square root of the
superconducting condensation energy per unit
volume. However, the precise value of $E_m$ could not be determined from the treatment of refs.\cite{prb03,edyn}.   The determination of the 
value of $E_m$ is one of the central results of this paper.  The interior charge density
$\rho_0$ depends on the sample dimensions (Eq. (2)) and becomes smaller as the sample size increases. 
 
We have also pointed out in earlier work that an electric field in the interior of a superconductor
should give rise to a macroscopic spin current\cite{sc0,sc}, and showed that in fact the 
microscopic Hamiltonian proposed by the theory of hole superconductivity\cite{prb89} favors the
existence of such a current\cite{sc}.  
 In very recent work we have put forth a more detailed picture of how the charge expulsion proposed to exist in superconductors occurs and how
 the spin current is generated\cite{sm}: namely,
 that electrons `expand' their wavefunction\cite{eh2} from being confined to a lattice spacing (as corresponds to a nearly filled band) to a much larger 
 (mesoscopic) extent.  
  As electrons move radially outward, the spin-orbit interaction gives rise to 
 azimuthal velocities resulting in the two members of the Cooper pair circulating in opposite directions
 {\it in an orbit of radius $2\lambda_L$,  which encloses precisely one flux quantum of spin-orbit flux}. This picture also provides
 for a $dynamical$ explanation of the Meissner effect\cite{sm}. It predicts that in the presence of an external magnetic field, 
 a charge {\it and a spin} current will circulate near the surface of a superconductor. In the absence of an external
 magnetic field, a pure spin current  will circulate, with a universal expression for the magnitude of the spin current speed at the surface 
 $v_{\sigma}^0  = \hbar /(4 m_e \lambda_L).$\cite{sm}

 For the charge current and charge density we derived their space and time dependence in Ref. \cite{edyn} (the charge current behavior predicted by our equations
 is the same as predicted by
 the conventional London equations). In this paper we derive the
 equations governing the space and time dependence of the spin current, and 
 show how the spontaneous spin current is related to the expelled negative  charge. Furthermore we provide a new interpretation of 
 type I and type II regimes in superconductors, and we explain how superconductors manage to conserve angular momentum. 
 We  also calculate the magnitude of the effects predicted   for various materials. Finally, we
 discuss the relation of our theory with  earlier pre-BCS work.
 
 We should point out  at the outset that the aim of this paper is limited to providing a macroscopic description of the electrodynamics of charge and spin
 currents in superconductors, and is $not$ to provide a full microscopic description of the superconducting state. Just like conventional BCS theory is consistent
 with conventional London electrodynamics, a full quantum-mechanical description of the problem should be 
 compatible with the electrodynamics discussed here   and will be the subject of future work.
  
 \section {spin current constitutive relation}
 
 For the charge current
 \beq
 \vec{J}=en_s \vec{v}_s
 \eeq
 with $n_s$ the superfluid density and $\vec{v}_s$ the superfluid $charge$ velocity, the governing equation is
 \beq
 \vec{J}(\vec{r})=-\frac{c}{4\pi\lambda_L^2}\vec{A}(\vec{r})
 \eeq
 with $\vec{A}$ the magnetic vector potential, related to the magnetic field $\vec{B}$ by
 \beq
 \vec{\nabla}\times\vec{A}=\vec{B}. 
 \eeq
 The London penetration depth
 is given by
 \beq
 \frac{1}{\lambda_L^2}=\frac{4\pi n_s e^2}{m_e c^2}
 \eeq
 and Eq. (5) is equivalent to
 \beq
 \vec{v}_s(\vec{r})=-\frac{e}{m_e c}\vec{A}(\vec{r})
 \eeq
 which can be understood as resulting from the 'rigidity' of the wave function that 'forces' $\vec{p}$ in the
 relation $\vec{p}=m_e\vec{v}_s+(e/c)\vec{A}$ to stay zero at all times and locations in a simply connected superconductor.
 
 We have shown in ref.\cite{sm} that the Meissner effect can be understood 'dynamically' by assuming
 that electrons move radially outward a distance $2\lambda_L$ in the presence of an $unscreened$ magnetic field
 $\vec{B}$ that gives rise to a vector potential
 \beq
 \vec{A}=\frac{\vec{B}\times\vec{r}}{2} ,
 \eeq
 and acquire through the action of the Lorentz force an azimuthal velocity given by Eq. (8). In a cylindrical geometry,
 the magnetic field at the surface equals the applied magnetic field $\vec{B}$, and the vector potential at the
 surface is given by
 \beq
 \vec{A}(R)=\lambda_L\vec{B}\times\hat{n}
 \eeq
 ($\hat{n}=\hat{r}$ denotes the direction normal to the surface) which coincides with Eq. (9) for $r=2\lambda_L$. Eq. (10) results from solving London's equation for 
 an infinitely long cylinder of radius $R>>\lambda_L$\cite{laue}, or from simply assuming that the magnetic field penetrates a distance $\lambda_L$ and using that
 \beq
 \oint \vec{A}\cdot d\vec{l}=\int \vec{B}\cdot d\vec{a}   \nonumber
 \eeq
 
 Similarly, we showed in ref.\cite{sm} that an 'effective' $unscreened$ magnetic field $\vec{B}_\sigma$ acts on the superfluid
 electrons arising from the spin-orbit interaction
 \beq
 \vec{B}_\sigma=2\pi n_s \vec{\mu}
 \eeq
 with
 \beq
 \vec{\mu}=\frac{e\hbar}{2m_e c} \vec{\sigma}
 \eeq
 the electron magnetic moment. Electrons acquire a spin current velocity\cite{sm}
 \beq
 \vec{v}_{\sigma}^0  =-\frac{\hbar}{4 m_e \lambda_L}\vec{\sigma}\times \hat{r}\equiv -v_\sigma^0\vec{\sigma}\times \hat{r}
 \label{key}
\eeq
through the action of the 'Lorentz' force from the field Eq. (11) in moving radially outward  a distance $2\lambda_L$. The spin-orbit
vector potential corresponding to 'magnetic field' Eq. (11) at radial distance $2\lambda_L$ is, from Eqs. (11) and (9)
\beq
\vec{A}_\sigma=2\pi n_s \lambda_L \vec{\mu}\times\hat{r}.
\eeq
Just like in the case of the real magnetic field, we assume that this is the value of the spin-orbit vector potential
{\it at the surface of the cylinder}.
Using Eq. (7), we can rewrite the spin-orbit magnetic field and vector potential  at the surface
Eqs. (11) and (14)  as
\bmath
 \beq
 \vec{B}_\sigma(R)= \frac{\hbar c}{4 e \lambda_L^2} \vec{\sigma}=\frac{m_e c }{e\lambda_L} v_\sigma^0\vec{\sigma}
 \eeq
 \beq
\vec{A}_\sigma(R)=\frac{\hbar c}{4 e \lambda_L} \vec{\sigma}\times\hat{n}=-\frac{m_e c}{e} \vec{v}_\sigma^0
\eeq
 \emath
respectively. 
Next, we need to deduce their behavior as we move away from the surface towards the interior.

We consider a cylindrical geometry throughout. We expect the spin current to flow within a London penetration
depth of the surface, just like the charge current, with the electron spin
parallel (antiparallel) to the cylinder axis and antiparallel (parallel) to its orbital angular momentum. 
The 'spontaneous'  current of the spin-$\sigma$ carriers is
\beq
\vec{J}_\sigma (\vec{r})=\frac{e n_s}{2} \vec{v}_{\sigma}(\vec{r}) .
\eeq
The total velocity of a carrier of spin $\vec{\sigma}$ in the presence of an external magnetic field is 
$\vec{v}_{\sigma,tot}=\vec{v}_{\sigma}+\vec{v}_s$. At the
surface ($r=R$) the spin current velocity $\vec{v}_\sigma(R)$  is given by Eq. (13), and the spin-orbit vector potential by Eq. (14) (or (15b)). 
The equation relating the spin current to the spin vector potential is, in analogy with Eq. (5)
\beq
\vec{J}_\sigma(\vec {r})=-\frac{c}{8\pi \lambda_L^2}\vec{A}_\sigma (\vec{r})
\eeq
and from Eqs. (16) and (17)
 \beq
 \vec{v}_{\sigma}(\vec{r})=-\frac{e}{m_e c}\vec{A}_\sigma(\vec{r})
 \eeq
 analogous to Eq. (8).
 
 As discussed in ref.\cite{sc}, the spin current exists due to the presence of an electric field $\vec{E}$ in the
 interior of the superconductor, that satisfies the differential equation\cite{edyn}
 \beq
\nabla^2(\vec{E}-\vec{E}_0)=\frac{1}{\lambda_L^2}(\vec{E}-\vec{E_0}),
\eeq
where $\vec{E}_0(\vec{r})$ is the electric field arising from the uniform 
positive charge distribution $\rho_0$ ($\vec{\nabla}\cdot\vec{E}_0=4\pi\rho_0$). $\vec{E}$ 
is maximum at a distance $\lambda_L$ from the surface and decays to zero at the surface
(for a cylindrical or spherical sample).
As discussed by Aharonov and Casher\cite{ac} and other authors\cite{other},
the spin-orbit interaction term in the non-relativistic limit of Dirac's equation for a 
particle with magnetic moment $\vec{\mu}$ in an electric field
can be written in terms of a spin-orbit vector potential proportional to $\vec{\mu}\times\vec{E}$.
We expect our spin-orbit vector potential to be related to the electric field in a similar fashion, and
 postulate  that the vector potential driving the spin current in Eq. (17) is determined by the
difference between the electric field near the surface $\vec{E}$ and the electric field in the deep interior $\vec{E}_0$
through the relation:
\beq
\vec{A}_\sigma(\vec{r}) =-\frac{m_e c v_\sigma^0}{e E_m}\vec{\sigma}\times (\vec{E}(\vec{r})-\vec{E_0}(\vec{r}))
\eeq
with $E_m$  related to the charge density near the surface $\rho_-$ by the charge conservation condition Eq. (3).

Eq. (20) reduces to Eq. (15b) for the spin-orbit vector potential at the surface, as is  seen by 
setting $\vec{E}(R)=0$ and $\vec{E}_0(R)=E_m\hat{n}$. From Eq. (18), the spin current velocity as function of
position is given by
\beq
 \vec{v}_{\sigma}(\vec{r})=\frac{v_\sigma^0}{E_m} \vec{\sigma}\times (\vec{E}(\vec{r})-\vec{E_0}(\vec{r})).
 \eeq
 and properly reduces to $\vec{v}_\sigma^0$  (Eq. (13)) at the surface, as required. The spin-orbit vector potential  and
 the spin current decay to zero as we move from the surface towards the interior of the superconductor
a distance $\lambda_L$, just like the corresponding charge quantities. 
Taking the curl of Eq. (20) ($\vec{\nabla}\times (\vec{\sigma}\times\vec{E})=\vec{\sigma}(\vec{\nabla}\cdot\vec{E})$) and using Gauss' law yields
\beq
\vec{B}_\sigma(\vec{r})=\vec{\nabla}\times\vec{A}_\sigma(\vec{r})=
-\frac{m_e c v_\sigma^0}{ e E_m} 4\pi (\rho(\vec{r}) - \rho_0) \vec{\sigma}
\eeq
At the surface $r=R$, $\rho(R) = \rho_-$, and if $R>>\lambda_L$ we can neglect $\rho_0$ in Eq. (22) and obtain:
 \beq
\vec{B}_\sigma(R)= 
-\frac{me c v_\sigma^0}{ e  E_m }4 \pi \rho_-  \vec{\sigma}
\eeq
which reduces to  the un-screened spin-orbit field Eq. (15a) upon replacing $\rho_-$ by its value given by 
Eq. (3).  The result Eq. (23)  followed from using Gauss' law and imposing 
the $global$ charge conservation condition  implicit in Eq. (3): that the positive charge expelled from the interior
that gives rise to the maximum electric field $E_m$ at  distance $\lambda_L$ from the surface is 
precisely the same as the extra negative charge $\rho_-$ residing in the surface layer of thickness $\lambda_L$.
Thus the coincidence of Eq. (23)  with Eq. (15a) {\it was not built in} and supports the consistency and validity of  our framework.

\section{The fourth component}
In ref.\cite{edyn} we showed that  the   electrodynamics in the charge sector, under the assumption that the magnetic vector potential obeys the Lorentz gauge,
could naturally be formulated in  a 4-dimensional covariant form in terms of the 4-dimensional charge current
and vector potential $J=(\vec{J},ic\rho)$, $A=(\vec{A},i\phi)$ ($\phi=$ electric potential, $\rho=$ charge density). 
The four-dimensional divergence of $J$ and of $A$ vanish due to charge conservation and the Lorentz gauge condition respectively.
Similarly we seek here the fourth component of a 4-dimensional spin vector
potential $A_\sigma=(\vec{A}_\sigma,i\phi_\sigma)$ by demanding that the four-dimensional divergence vanishes:
\beq
Div A_\sigma=\vec{\nabla}\cdot \vec{A}_\sigma+\frac{1}{c}\frac{\partial}{\partial t}\phi_\sigma=0.
  \eeq
  Using the relation
  \beq
  \vec{\nabla}\cdot(\vec{\sigma}\times\vec{E})=-\vec{\sigma}\cdot(\vec{\nabla}\times \vec{E}) \nonumber
  \eeq
and Faraday's law we obtain  from Eqs. (20) and (24)
  \beq
  \phi_\sigma=\frac{m_e c v_\sigma^0}{e E_m}\vec{\sigma}\cdot\vec{B}
  \eeq
  and correspondingly we   add the fourth component $\rho_\sigma$ to form the four-current for carriers of spin $\sigma$,  $J_\sigma=(\vec{J}_\sigma,ic\rho_\sigma)$, satisfying the equation (cf. Eq. (17))
     \bmath
       \beq
  \rho_\sigma=-\frac{m_e c v_\sigma^0}{8\pi e E_m \lambda_L^2} \vec{\sigma}\cdot\vec{B}.
  \eeq  
This equation can be rewritten as
  \beq
  \rho_\sigma=-\frac{en_s}{2E_m} \frac{v_\sigma^0}{c}  \vec{\sigma}\cdot\vec{B} 
    \eeq
  or as 
    \beq
     \rho_\sigma=-\frac{n_s}{4\lambda_L E_m}  \vec{\mu}\cdot\vec{B}
\eeq
  \emath
and implies that in the presence of an applied magnetic field there is excess negative charge corresponding
  to spin direction antiparallell to the applied field $\vec{B}$. This corresponds to the electrons near the surface
  that $increase$ their velocity when $\vec{B}$ is applied.
  Correspondingly, the charge density of the spin component that decreases its velocity when  $\vec{B}$ is applied decreases.
  Note that $\vec{J}_\sigma$ and $\rho_\sigma$ are related by the continuity equation $Div J_\sigma=0$:
  \beq
  \vec{\nabla}\cdot\vec{J}_\sigma+\frac{\partial \rho_\sigma}{\partial t}=0
  \eeq
  When an external magnetic field is applied, $\vec{J}_\sigma$ near the surface acquires a divergence through the induced $\vec{\nabla}\times\vec{E}$,
  which creates the spin imbalance through Eq. (27).
   
   \section{Magnitude of expelled negative charge and internal electric field}
   
   An applied magnetic field generates a charge current near the surface, and according to Eq. (26) it also generates a charge density.
   It is natural to conclude that this charge density arises due to the changed carrier velocity. From Eqs. (8) and (10), the
   magnetic field expressed in terms of the superfluid charge velocity at the surface is
   \beq
   \vec{B}=\frac{m_e c}{e\lambda_L}\vec{v}_s(R)\times\hat{n}
   \eeq
   and the induced charge density Eq. (26a) is
   \beq
   \rho_\sigma=-(\frac{m_e c}{e\lambda_L})^2\frac{v_\sigma^0}{8\pi E_m}  \vec{\sigma} \cdot(\vec{v}_s(R)\times\hat{n})
   \eeq
   We argue that it is natural to conclude that the expelled charge density $\rho_-$ will obey the same relation with respect to the
   spin current velocity $v_\sigma$  that $\rho_\sigma$ bears to the charge current velocity $v_s$ (Eq. (29)), 
   with an extra factor of $2$ because of the two spin components contributing to the spin current.
   Hence we postulate that
      \beq
   \rho_-=-(\frac{m_e c}{e\lambda_L})^2\frac{v_\sigma^0}{4\pi E_m} \vec{\sigma} \cdot(\vec{v}_\sigma^0\times\hat{n})
   \eeq
   or, substituting for $\rho_-$ using Eq. (3)
   \beq
   E_m=(\frac{m_e c}{e\lambda_L} v_\sigma^0)^2\frac{1}{E_m}
   \eeq
   with solution
   \beq
   E_m=-\frac{m_e c}{e\lambda_L}v_\sigma^0
   \eeq
   or, substituting for the spin current velocity from Eq. (13)
      \beq
   E_m=-\frac{\hbar c}{4e\lambda_L^2}=\frac{\phi_0}{4\pi \lambda_L^2}=2\pi n_s \mu_B
   \eeq
with $\phi_0=hc/2|e|$ the flux quantum and $\mu_B$ the Bohr magneton. Hence {\it the magnitude of the internal electric field near the surface is the same as that of the
spin-orbit effective magnetic field}  (Eq. (11) or (15a)).

The value of the expelled charge density near the surface, $\rho_-$,  follows from Eqs. (3) and (32):
\beq
\rho_-=n_s e \frac{v_\sigma^0}{c}.
\eeq
which can also be written as, using Eq. (13)
\beq
\rho_-=\frac{n_s e\hbar}{4\lambda_L m_e c}=-\frac{n_s}{2\lambda_L} \mu_B
\eeq
 or, using Eq. (7), as
\beq
\rho_-=\frac{\hbar c}{16\pi e \lambda_L^3}
\eeq

The charge density induced by the external magnetic field 
can now be written, substituting in Eq. (26a) $E_m$ by its value Eq. (32) 
\bmath
\beq
\rho_\sigma=\frac{1}{8\pi \lambda_L}\vec{\sigma}\cdot\vec{B}
\eeq
or, substituting for $E_m$ in Eq. 26(b)
\beq
\rho_\sigma =\frac{n_s e}{2c}\vec{\sigma}\cdot(\vec{v}_s\times\hat{n}) \eeq
\emath
as one would expect by analogy with Eq. (34). Thus the total charge density for spin $\sigma$ is simply
\beq
\rho_{\sigma,tot}=\frac{n_s e}{2} (\frac{v_\sigma^0+\sigma v_s}{c})=\frac{n_s e}{2} \frac{\sigma v_{\sigma,tot}}{c}
\eeq
with $v_{\sigma,tot}=v_s+\sigma v_\sigma^0$,    $\sigma=+/-1$ for spin parallel/antiparallel to the applied magnetic field, and
\beq
\rho_-=\rho_{\sigma=1,tot}+\rho_{\sigma=-1,tot}
\eeq
independent of whether or not an external magnetic field is applied.

Equation (5) for the charge current can be written in terms of the expelled charge density  $\rho_-$ using Eq. (3) as
\beq
\vec{J}=\rho_- c \frac{ \vec{B} \times \hat{n}}{E_m}
\eeq
where we have used Eq. (10) for the vector potential. Eq. (40)  has the following interpretation: rather than the entire superfluid moving
with speed $v_s$ (Eq. (4)), we may think of the charge current as being carried solely by the $excess$ negative charge density $\rho_-$, moving with velocity
\beq
\vec{v}_{\rho_-}=c\vec{B} \times \hat{n}/E_m
\eeq
(Note that the total charge density $\rho_-$ is not changed by the applied magnetic field since $\rho_\sigma=-\rho_{-\sigma}$).
The speed Eq. (41) would exceed the speed of light if the applied magnetic field would exceed $E_m$. However, the value we deduced for
$E_m$ Eq. (33) is just right to prevent this from happening, since it is of the order of the lower critical field for a type II superconductor\cite{tinkham}:
\beq
H_{c1}= \frac{\phi_0}{4\pi \lambda_L^2} ln\kappa
\eeq
with $\kappa$ the Ginzburg Landau parameter. Furthermore, we argued in Ref.\cite{sm} that superconductivity is destroyed when one of the
components of the spin current is stopped by the applied magnetic field, which corresponds to a magnetic field of magnitude $E_m$.
Thus, Eq. (41) can be understood as meaning that superconductivity is destroyed when the speed of the 
excess charge carriers near the surface ($\rho_-$) reaches the speed of light.

An alternative and perhaps even more remarkable interpretation follows from writing the charge current Eq. (4) solely in terms of the 
charge density induced by the magnetic field Eq. (37), as:
\beq
\vec{J}=en_s\vec{v}_s=c(\rho_{\sigma=-1}-\rho_{\sigma=1})\hat{v}_s.
\eeq
Eq. (43) can be read as meaning that the charge current induced by the magnetic field is carried solely by the induced charge carriers moving
at the speed of light. Within this interpretation, superconductivity is destroyed when the induced charge density of the carriers that are
slowed down by the applied magnetic field (carriers with spin $parallel$ to the magnetic field) completely depletes the component of $\rho_-$ with that spin orientation.

In a type I superconductor, superconductivity is destroyed when the applied magnetic field reaches the 
thermodynamic critical field $H_c$, which is smaller than Eq. (33) 
($H_c\sim H_{c1}\times\lambda_L/\xi_0$, with $\xi_0$ the Pippard/BCS coherence length). However,
in that case the response to an applied magnetic field is non-local since the magnetic field varies strongly over a
coherence length, so these considerations
need to be modified. We will argue in a later section that for a type I superconductor $E_m=H_c$ rather than
Eq. (33). The expelled charge density $\rho_-$ is still given by Eq. (3), which implies that the interpretation of Eq. (41) is 
the same as discussed above for type II superconductors. Fig. 1 shows schematically the spin current and charge distribution in a cross section of a cylindrical sample.

     \begin{figure}
\resizebox{8.0cm}{!}{\includegraphics[width=7cm]{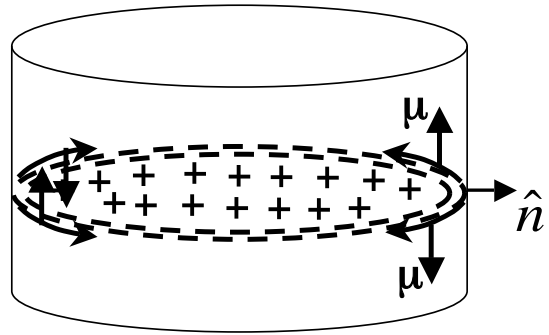}}
\caption{Charge distribution and spin current in a cross section of a cylindrical sample. The charge density is negative within $\lambda_L$ of the surface and positive in the interior.
The spin current circulates where the charge density is negative. At the surface, the velocity is given by Eq. (13). The horizontal arrows indicate the direction of motion of the
electrons with magnetic moment in the direction indicated by the vertical arrows.}
\label{atom5}
\end{figure}

The electrostatic energy cost per unit volume for a surface charge density $\rho_-$ expelled from a volume $V$
  is given by (since  $\rho_0$ is uniform in the interior)
\beq
u=\frac{1}{V}\int \frac{E^2}{8\pi} dv=\frac{1}{2}\frac{E_m^2}{8\pi}
\eeq
for the case of a cylinder
and 
\beq
u=\frac{3}{5}\frac{E_m^2}{8\pi}
\eeq
 for a sphere, so   the electrostatic energy density is smaller than the magnetic energy density
associated with the thermodynamic critical  field $H_c^2/8\pi$ for both type I and type II superconductors.

It is interesting to note that Eq. (34) together with the charge neutrality condition Eq. (3) and Eq. (7) lead to
\beq
v_\sigma^0=\frac{e}{m_ec}\lambda_L E_m
\eeq
of the same form as the relation between charge velocity and magnetic field (Eq. (28)). The kinetic energy density of the spin current is, from Eqs. (46) and (7)
\beq
\frac{1}{2} m_e (v_\sigma^0)^2 n_s=\frac{E_m^2}{8\pi}
\eeq
hence it equals the electrostatic energy density near the surface. As is well known, the same relation exists between the kinetic energy density associated with the
charge current and the magnetic energy density near the surface. Hence the total kinetic energy density in the presence of an applied $B$ and the
associated charge current is
\beq
\frac{1}{2} m_e(v_{\sigma=1,tot}^2+v_{\sigma=-1,tot}^2)\frac{n_s}{2}=\frac{E_m^2+B^2}{8\pi}
\eeq
The charge current screens the external magnetic field so it does not get $into$ the superconductor. The spin current screens the internal electric field so it does not
leak $out$ $of$ the superconductor. 

\section{Spin current spatial distribution}
Having determined the value of the internal electric field $E_m$, we return to the formulation of the equations governing the spin current. 
We will slightly generalize the equations in Sect. II. As the equation relating spin current to spin vector potential we take instead of Eq. (17)
\beq
\vec{J}_\sigma(\vec {r})-\vec{J}_{\sigma 0}=
-\frac{c}{8\pi \lambda_L^2}(\vec{A}_\sigma (\vec{r})-\vec{A}_{\sigma 0} (\vec{r}))
\eeq
with
\bmath
\beq
\vec{A}_\sigma(\vec{r}) =\lambda_L \vec{\sigma}\times \vec{E}(\vec{r})
\eeq
\beq
\vec{A}_{\sigma 0}(\vec{r}) =\lambda_L \vec{\sigma}\times \vec{E}_0(\vec{r})
\eeq
\emath
and
\bmath
\beq
\vec{J}_\sigma(\vec{r})=\frac{en_s}{2}\vec{v}_\sigma (\vec{r})
\eeq
\beq
\vec{J}_{\sigma 0} =\frac{en_s}{2}\vec{v}_{\sigma 0} 
\eeq
\emath
The coefficient $\lambda_L$  in Eq. (50) results from replacing in Eq. (20) the value found for $E_m$, Eq. (32).  Note the similarity between  Eq. (50a) and Eq. (10), 
which  was not ``built in''.
Aside from an inconsequential redefinition of $\vec{A}_\sigma$ ($\vec{A}_\sigma$ in Eq. (17) is ($\vec{A}_\sigma-\vec{A}_{\sigma 0})$ in Eq. (49)), Eqs. (49)-(51) differ from
Eqs. (17) and (20) in that we allow for the possibility of a constant current $\vec{J}_{\sigma 0}$ deep in the interior of the
superconductor, where $\vec{E}=\vec{E}_0$. From Eq. (49) we obtain using Eqs. (50) and (51):
\beq
 \vec{v}_{\sigma}(\vec{r})-\vec{v}_{\sigma 0}=-\frac{c}{4 \pi e n_s \lambda_L}    \vec{\sigma}\times (\vec{E}(\vec{r})-\vec{E_0}(\vec{r})) 
 \eeq
 and taking the curl
  \beq
 \vec{\nabla}\times (\vec{v}_{\sigma}(\vec{r})-\vec{v}_{\sigma 0})=-\frac{c}{en_s}\frac{\rho(\vec{r})-\rho_0}{\lambda_L} \vec{\sigma}  .
 \eeq
 Eq. (53) is approximately satisfied by
 \bmath
    \beq
    \vec{v}_{\sigma}(\vec{r})=-\frac{c}{en_s}\rho(\vec{r}) \vec{\sigma}\times\hat{r}
    \eeq
    with
        \beq
    \vec{v}_{\sigma 0}=\vec{v}_\sigma(r<<R)=-\frac{c}{en_s}\rho_0 \vec{\sigma}\times\hat{r}
    \eeq
    \emath
    since using these expressions we obtain
       \beq
     \vec{\nabla}\times (\vec{v}_{\sigma}(\vec{r})-\vec{v}_{\sigma 0})=
     -\frac{c}{en_s}\frac{\partial \rho(\vec{r})}{\partial r}\vec {\sigma}
     -\frac{c}{en_s}\frac{\rho(\vec{r})-\rho_0}{r}\vec {\sigma}
     \eeq
     The second term in Eq. (55) is zero in the deep interior and is smaller than the first term by a factor $\lambda_L/R$ near the surface,
     and the first term in Eq. (55) is approximately the same as Eq. (53).
     
     Eq. (54) generalizes the relation found between charge density and velocity of the spin current carriers near the surface, Eq. (37), to the
     entire volume. The form Eq. (54) is only valid in the absence of applied magnetic field, when only the pure spin current exists
     ($\vec{v}_{\sigma=+1}=-\vec{v}_{\sigma=-1})$. In the presence of both charge and spin current, we write instead of Eq. (54) 
     \beq
      \vec{v}_{\sigma}(\vec{r})=-\frac{2c}{en_s}\rho_\sigma(\vec{r}) \vec{\sigma}\times\hat{r}
    \eeq
 where (modifying the convention used in Sect. IV) we denote by $\rho_\sigma$ the $total$ charge density for spin $\sigma$, produced by
     both the pure spin current and any superimposed charge current induced by an external magnetic field
     (this was called $\rho_{\sigma,tot}$ in Sect. IV), and 
     $v_\sigma$ denotes what was called $v_{\sigma,tot}$ in Sect. IV Eq. (38).
     
     In Sect. II we had argued that the spin current should die down as we move beyond a London penetration depth of the surface towards the interior,
     by analogy with the behavior of the charge current, and formulated the equations accordingly. They corresponded to taking
     $\vec{J}_{\sigma 0}=0$ in Eq. (49). However, the finding of a general relation between charge density and current in Sect. IV led us
     to include here the   $\vec{J}_{\sigma 0}$ term in the theory, representing a counterflowing spin current in the interior induced by the 
     charge density $\rho_0$, smaller than the spin current near the surface by a factor $\lambda_L/R$.  
     
     It should be noted however that the addition of the constant term $\vec{J}_{\sigma 0}$ leads to a   singularity in the vorticity in the deep interior: namely,
     for $r<<R$
     \beq
     \vec{\nabla} \times \vec{v}_\sigma(\vec{r})=-\frac{c}{en_s}\frac{\rho_0}{r}\vec{\sigma}
     \eeq
     diverges as $r \rightarrow 0$. We return to this point in Sect. XII.
     
     The spin current Eq. (51) expressed in terms of the  carrier's  velocity Eq. (56) takes the form
          \beq
     \vec{J}_\sigma(\vec{r})=-c\rho_\sigma(\vec{r})\vec{\sigma}\times\hat{r}
     \eeq
     and denotes the total current carried by the carriers of spin $\sigma$, including the contribution from the spontaneous pure spin current
     and from the charge current induced by an applied magnetic field, if any. Eq. (58) is consistent with Eq. (43) deduced for the 
     charge current only. 
          
          \section{Spill-over}
          The addition of the constant term $\vec{J}_{\sigma 0}$ to the spin current constitutive relation has another interesting consequence.
          Consider the equation for the spin current
          \beq
          \vec{J}_\sigma (\vec{r})-\vec{J}_{\sigma 0}=-\frac{c}{8\pi \lambda_L} \vec{\sigma}\times[\vec{E}(\vec{r})-\vec{E}_0(\vec{r})]
          \eeq
          We argued earlier that at the surface $\vec{E}=0$, $\vec{E}_0=E_m\hat{r}$, and the spin current velocity is given by the universal
          form Eq. (13). However, these arguments were based on the constitutive relation Eq. (17), that did not include the
          $\vec{J}_{\sigma 0}$ term. Eq. (59) no longer satisfies this condition. How is this inconsistency resolved?
          
          Consider a cylinder of radius $R$. The electric field $\vec{E}_0$ at the surface is
 \beq
 \vec{E}_0(R)=2\pi R \rho_0 \hat{n}
 \eeq
 Assuming the electric field $\vec{E}$ vanishes at the surface, Eq. (59) yields
 \beq
 \frac{e n_s}{2} \vec{v}_\sigma(R)-\frac{c\rho_0}{2}\vec{\sigma}\times\hat{n}=\frac{c}{4\lambda_L} R \rho_0 (\vec{\sigma}\times\hat{n})
 \eeq
 Bringing the second term to the right side, we have
 \beq
 \frac{e n_s}{2} \vec{v}_\sigma(R)=\frac{c}{4\lambda_L}  \rho_0 (R+2\lambda_L) (\vec{\sigma}\times\hat{n})
 \eeq
 which is satisfied {\it if we replace $R$ by $R+2\lambda_L$ in Eq. (2a)}, i.e. if the 'effective' radius of the cylinder is assumed to be
 \beq
 R_{eff}=R+2\lambda_L
 \eeq
instead of $R$, which indicates that the negative superfluid 'spills over' a distance $2\lambda_L$ beyond the radius of the cylinder.

To obtain the magnitude of the charge that spilled over, note that at $r=R$ the electric field $E(R)$ will no longer be zero. To satisfy Eq. (56)
with the spin current velocity  Eq. (13) and $E_0=E_m$ (Eq. (33)) requires an electric field at $R$ pointing outward, of magnitude
\beq
E(R)=4\pi \lambda_L \rho_0
\eeq
which corresponds to a sheet of surface charge density
\beq
\sigma_{spill}=-\lambda_L \rho_0
\eeq
having spilled out beyond the radius $R$. Being spread out over a radial distance $2\lambda_L$ beyond the surface, it corresponds to a spill-over volume
charge density $\sigma_{spill}/(2\lambda_L)$, i.e.
\beq
\rho_{spill}=-\frac{\rho_0}{2} .
\eeq

Note also that because of this spill-over effect, the actual negative charge density near the surface of the superconductor is not $\rho_-$,
rather it is $\rho_- + \rho_0$. Hence upon taking the curl of Eq. (20) we need to use
\beq
\vec{\nabla}\cdot(\vec{E}-\vec{E}_0)_{r=R}=4\pi((\rho_-+\rho_0)-\rho_0)=4\pi \rho_-
\eeq
This resolves the small discrepancy that we had found in Sect. II between the expressions for the unscreened spin-orbit
magnetic field Eqs. (22) and Eq. (15a).

Note also that this scenario is consistent with the microscopic picture put forth in ref.\cite{sm}, that   electrons expand their wavefunctions
effective radius 
from a lattice spacing to circular orbits of radius $2\lambda_L$ when a metal goes superconducting. For pairs that are right at the surface
in the normal state, these orbits will expand to a distance $2\lambda_L$ beyond the body's surface in the superconducting state. 

For a sphere of radius $R$, the total spilled charge is $4\pi R^2 \sigma_{spill}$, which using Eq. (2b) and (34) corresponds to a number
of spilled electrons
\beq
N_{spill}=\frac{3\hbar c}{16 \pi e^2 \lambda_L} R
\eeq
For example, for $\lambda_L=400\AA$ and $R=1cm$ this yields $N_{spill}=2\times 10^6$. We had  already predicted earlier such a spill-over
of negative charge beyond the surface of superconductors based on quite different arguments\cite{sc0}.

    \section{Spin current electrodynamics}
    We can now simply extend these equations to describe the full space and time dependence of the currents and charge densities.
    In terms of the four-vector for spin-component $\sigma$ the total current with or without an applied magnetic field is given by
    \beq
J_\sigma(\vec {r},t)-J_{\sigma 0}=-\frac{c}{8\pi \lambda_L^2}(A_\sigma (\vec{r},t)-A_{\sigma 0} (\vec{r}))
\label{constitutive}
\eeq
with
\bmath
\beq
 J_\sigma(\vec{r},t)=(\vec{J}_\sigma(\vec{r},t),ic\rho_\sigma(\vec{r},t))
\eeq
\beq
 J_{\sigma 0} =(\vec{J}_{\sigma 0},ic\rho_{\sigma 0})
\eeq
\emath
and
\bmath
\beq
A_\sigma(\vec{r},t)=(\vec{A}_\sigma(\vec{r},t),i\phi_\sigma(\vec{r},t))
\eeq
\beq
A_{\sigma 0}(\vec{r})=(\vec{A}_{\sigma 0}(\vec{r}),i\phi_{\sigma 0}(\vec{r}))
\eeq
\emath
with
\bmath
\beq
\vec{A}_\sigma(\vec{r},t) =\lambda_L \vec{\sigma}\times \vec{E}(\vec{r},t)+\vec{A}(\vec{r},t)
\eeq
\beq
\vec{A}_{\sigma 0}(\vec{r}) =\lambda_L \vec{\sigma}\times \vec{E}_0(\vec{r})
\eeq
\emath
and
\bmath
\beq
  \phi_\sigma(\vec{r},t)=-\lambda_L\vec{\sigma}\cdot\vec{B}(\vec{r},t)+\phi(\vec{r},t)
  \eeq
  \beq
  \phi_{\sigma 0}(\vec{r})=\phi_0(\vec{r})
  \eeq
  \emath
  Here, $\vec{A}$ and $\phi$ are the magnetic vector potential and electric potential. $\vec{E}_0$ and $\phi_0$ are the electrostatic field and
  potential for a uniform charge density $\rho_0$ throughout the material. The spatial component of Eq. (\ref{constitutive}) is
  \bmath
  \beq
   \vec{J}_\sigma(\vec{r},t)- \vec{J}_{\sigma 0}=-\frac{c}{8\pi \lambda_L^2}(\lambda_L \vec{\sigma}\times \vec{E}(\vec{r},t)+\vec{A}(\vec{r},t))
  \eeq
  and the fourth component is 
    \beq
  \rho_\sigma(\vec{r},t)-\rho_{\sigma 0}=\frac{1}{8\pi \lambda_L} \vec{\sigma}\cdot \vec{B}(\vec{r},t)-
  \frac{1}{8\pi \lambda_L^2} 
  (\phi(\vec{r},t)-\phi_0(\vec{r}))
  \eeq
  \emath
   The continuity equation sets the four-dimensional divergence of the four-vector $J_\sigma$ equal to zero
  \beq
  Div J_\sigma=0
  \eeq
  with the fourth component of the divergence operator given by $\partial/\partial(ict)$. The Lorentz gauge condition
  \beq
  \vec{\nabla}\cdot \vec{A}+\frac{1}{c}\frac{\partial \phi}{\partial t}=0
  \eeq
together with Faraday's law $\vec{\nabla}\times\vec{E}=(-1/c)\partial\vec{B}/\partial t$ ensures that the four-divergence of $A_\sigma$ vanishes, consistent with Eq. (69).
That $A_\sigma$ is a four-vector can be seen from the fact that it can be written as
\beq
(A_\sigma)_\alpha=\frac{i\lambda_L}{2} \epsilon_{\alpha \beta \gamma \delta}\sigma_\beta F_{\gamma \delta}+A_\alpha
\eeq
with $F_{\gamma \delta}$ the electromagnetic field tensor, $A_\alpha$ the usual electromagnetic four-vector potential
obeying the Lorentz gauge condition, and $ \epsilon_{\alpha \beta \gamma \delta}=+1$ ($-1$) for even (odd) permutations of $1234$ and zero otherwise.
  
  The current 4-vectors are given in terms of the velocity of the superfluid charge density per spin $en_s/2$, the velocity for each
  spin component $v_\sigma$, and the (excess) charge density $\rho_\sigma$  as
  \bmath
  \beq
  J_\sigma (\vec{r},t)=(\frac{en_s}{2}\vec{v}_\sigma(\vec{r},t),ic\rho_\sigma(\vec{r},t))
  \eeq
    \beq
  J_{\sigma 0}=(\frac{en_s}{2}\vec{v}_{\sigma 0},ic\rho_{\sigma 0})
  \eeq
    \emath
  with $\vec{v}_{\sigma 0}$ given by Eq. (51b) and $\rho_{\sigma 0}=\rho_0/2$.   
  Using the relation (56), they can be written in the 
  remarkable form
    \bmath
  \beq
  J_\sigma (\vec{r},t)=\rho_\sigma(\vec{r},t)c(-\vec{\sigma}\times\hat{r},i) 
  \eeq
    \beq
  J_{\sigma 0}=\frac{\rho_0 c}{2}(-\vec{\sigma}\times\hat{r},i)  
   \eeq
    \emath
  which says that the  supercurrent density (charge or spin) at any point in the superconductor  can be understood as arising from the
  excess local charge density moving at the speed of light.

    The differential equations determining the behavior of all quantities are
    \bmath
\beq
\Box^2 ( A_\sigma-A_{\sigma 0})=\frac{1}{\lambda_L^2}(A_\sigma -A_{\sigma 0})
\eeq
\beq
\Box^2 ( J_\sigma-J_{\sigma 0})=\frac{1}{\lambda_L^2}  (J_\sigma-J_{\sigma 0}) .
\eeq
\emath
with 
\beq
\Box^2=\nabla^2-\frac{1}{c^2}\frac{\partial^2}{\partial t^2}
\eeq
and $J_\sigma$ is given in terms of  $A_\sigma$ by Eq. (69). The equations for the charge sector only are simply obtained by defining the charge four-current and charge four-potential
\bmath
\beq
J_c=J_{\sigma=+1}+J_{\sigma=-1}
\eeq
\beq
A_c=(A_{\sigma=+1}+A_{\sigma=-1})/2
\eeq
\emath
and similarly for $J_{c0}$ and $A_{c0}$, 
and they obey
   \beq
J_c(\vec {r},t)-J_{c 0}=-\frac{c}{4\pi \lambda_L^2}(A_c (\vec{r},t)-A_{c 0} (\vec{r}))
\eeq
and Eq. (83), which of course coincide with the equations derived in ref.\cite{edyn} for the charge sector.

   \section{Spin current electrostatics}

In the absence of time-dependence, the electric field in the interior of the superconductor is described
by the differential equation Eq. (19),  and the charge density obeys the differential equation Eq. (1). They are  related by the equation
\beq
\vec{E}(\vec{r})-\vec{E}_0(\vec{r})=4\pi \lambda_L^2\vec{\nabla}\rho(\vec{r})
\eeq
For a cylinder, the solution to these equations is given in terms of modified Bessel functions $I_n(z)$ (Bessel functions of imaginary argument) as\cite{laue}
\bmath
\beq
\rho(r)=\rho_0(1-\frac{i R}{2\lambda_L} \frac{I_0(ir/\lambda_L)}{I_1(iR/\lambda_L)})
\eeq
\beq
\vec{E}(r)=2\pi\rho_0 \vec{r}[1-\frac{R}{r} \frac{I_1(ir/\lambda_L)}{I_1(iR/\lambda_L)}]
\eeq
\emath
 for a sphere
\bmath
\beq
\rho(r)=\rho_0(1-\frac{1}{3}\frac{R^3}{\lambda_L^2 r}\frac{sinh(r/\lambda_L)}{f(R/ \lambda_L)})
\eeq
\beq
\vec{E}(r)=\frac{4}{3}\pi\rho_0\vec{r}[1-\frac{R^3}{r^3}\frac{f(r/\lambda_L)}{f(R/\lambda_L)}]
\eeq
\emath
with $f(x)=x coshx-sinhx$,
and for a plane
\bmath
\beq
\rho(r)=\rho_0(1-\frac{R}{\lambda_L}\frac{cosh(r/\lambda_L)}{sinh(R/\lambda_L)})
\eeq
\beq
\vec{E}(r)=4\pi \rho_0 \vec{r}[1-\frac{R}{r}\frac{sinh(r/\lambda_L)}{sinh(R/\lambda_L)}]
\eeq
\emath

For $R>>\lambda_L$ they take the following forms: for the cylinder
\bmath
\beq
\rho(r)=\rho_0(1-\frac{R^{3/2}}{2\lambda_Lr^{1/2}}e^{-(R-r)/\lambda_L})
\eeq
\beq
\vec{E}(r)=2\pi\rho_0 \vec{r}[1-\frac{R^{3/2}}{r^{3/2}}e^{-(R-r)/\lambda_L}]
\eeq
\emath
and $\vec{E}_0=2\pi\rho_0 \vec{r}$; 
 for the sphere
\bmath
\beq
\rho(r)=\rho_0(1-\frac{R^{2}}{3\lambda_Lr}e^{-(R-r)/\lambda_L})
\eeq
\beq
\vec{E}(r)=\frac{4}{3}\pi\rho_0\vec{r}[1-\frac{R^{2}}{r^2}e^{-(R-r)/\lambda_L}]
\eeq
\emath
and $\vec{E}_0=(4 \pi/3)\rho_0\vec{r}$, and for the plane
\bmath
\beq
\rho(r)=\rho_0(1-\frac{R}{\lambda_L}e^{-(R-r)/\lambda_L})
\eeq
\beq
  \vec{E}(r)=4\pi\rho_0\vec{r}[1-\frac{R}{r}e^{-(R-r)/\lambda_L}]
  \eeq
  \emath
and $\vec{E}_0=4\pi \rho_0 \vec{r}$.

The spin current is given in terms of the electric field by
\beq
\vec{J}_\sigma(\vec{r})=-\frac{c}{8\pi  \lambda_L}\vec{\sigma}\times(\vec{E}-\vec{E}_0)+\vec{J}_{\sigma 0}
\eeq
with
\beq
\vec{J}_{\sigma 0}=-\frac{c}{2}\rho_0\vec{\sigma}\times\hat{r}
\eeq
 For the cylinder, Eq. (91) yields
 \beq
\vec{J}_\sigma(\vec{r})=\frac{\rho_0 c}{2}[\frac{R}{2\lambda_L} e^{-(R-r)/\lambda_L}-1]\vec{\sigma}\times\hat{r}
\eeq
and a similar expression for the sphere. Both can   be written as
 \beq
\vec{J}_\sigma(\vec{r})=-\frac{c}{2}[\rho_-e^{-(R-r)/\lambda_L}+\rho_0]\vec{\sigma}\times\hat{r} = 
\eeq
or
\beq
\vec{J}_\sigma(\vec{r})= \vec{J}_\sigma (R) e^{-(R-r)/\lambda_L}+\vec{J}_{\sigma 0}
\eeq
The magnitude of the spin current near the surface is
\beq
J_\sigma(R)=1.02\times 10^8 \frac{Amp}{cm^2}\times (\frac{400 \AA}{\lambda_L(\AA)}^3)
\eeq
and in the deep interior
\beq
J_{\sigma 0}=\frac{2\lambda_L}{R}J_\sigma(R)   .
\eeq

The direction of the spin vector $\vec{\sigma}$ is arbitrary in the spherical geometry. If an external magnetic field is applied in the $z$ direction, 
a charge current develops according to Eq. (5) in the azimuthal direction and defines the spin quantization axis. 
Then the up and down spin components of the
total current are simply obtained by using the $z$ axis as the spin quantization axis and adding the charge and spin current contributions.
Similarly, a magnetic field applied parallel to a planar surface will polarize the spin current with the spin quantization axis parallel to the applied field.

If an external electric field $\vec{E}_{ext}$ is applied, Eq. (19) remains valid but the boundary condition changes\cite{taopaper}. In a spherical or
cylindrical geometry there is no external electric  field in the absence of applied electric field, hence we have at the surface
\beq
\vec{E}=\vec{E}_{ext}
\eeq
Assuming there are no 'normal electrons', the electric field is continuous at the surface and penetrates a distance $\lambda_L$\cite{edyn}. So right at the
surface we have
\beq
\vec{J}_\sigma(\vec{r})=-\frac{c}{8\pi  \lambda_L}\vec{\sigma}\times(\vec{E}_{ext}-\vec{E}_0)+\vec{J}_{\sigma 0}
\eeq
Consequently, an external electric field that points $in$ (antiparallel to $\vec{E}_0$) will enhance the spin current, and one that 
points $out$ will suppress it. If the magnitude of the electric field applied is larger than the critical field $E_m$ (eq. 33) it will 
completely suppress the spin current and hence destroy superconductivity.

 If an electric current circulates along the axial direction of the cylinder, one expects that in the absence of applied electric field that
 above a critical current
 \beq
 J_{c1}=\frac{c}{4\pi} \frac{H_{c1}}{\lambda_L}
 \eeq
 vortices start to penetrate the superconductor\cite{tinkham}. If an electric field normal to the surface is applied, this expression will be modified to
 \beq
 J_{c1}=\frac{c}{4\pi} \frac{H_{c1}\pm E_{ext}}{\lambda_L}
 \eeq
 with the $+$ ($-$) sign corresponding to electric field pointing into (out of) the cylinder. Thus,  for a cylinder placed between the
 plates of a capacitor   the flux penetration for a given current flowing along the cylinder should be affected by the electric field of the capacitor and depend on the
 position in the surface relative to the capacitor plates: the flux penetration should be larger for the region near 
 the $negative$ capacitor plate.
 
  \section{Origin of the spin current}
  As discussed in Ref.\cite{sm}, the physics that we are proposing as underlying both the generation of spin current and the Meissner effect in superconductors is an
 $expansion$ of the electronic wavefunction\cite{eh2}, from a lattice spacing
   to a radius $2\lambda_L$, as the system enters the
 superconducting state. In the presence of a magnetic field, the azimuthal Lorentz force on radially outgoing electrons gives rise to the   charge current that 'expels' the magnetic field.
 An azimuthal spin-orbit force originating in the positive background\cite{fm} gives rise to the spin current, whether or not an external
 magnetic field is present.

 The fact that a magnetic moment propagating in a periodic lattice experiences a net transverse force was proposed in Ref. \cite{fm} as an explanation for the anomalous
 Hall effect in ferromagnetic metals. An electron with magnetic moment $\vec{\mu}$ propagating with velocity $\vec{v}$ is equivalent to an electric dipole\cite{dipole,dipole2}
 \beq
 \vec{p}=\gamma\frac{\vec{v}}{c}\times\vec{\mu}
 \eeq
 with $\gamma=(1-v^2/c^2)^{-1}\sim1$ for non-relativistic speeds. An electric dipole in an electric field experiences a
  force $\vec{F}=\vec {\nabla}(\vec{p}\cdot\vec{E})$ and a torque $\vec{\tau}=\vec{p}\times\vec{E}$.  As shown in Fig. 2(a), for a discrete positive charge distribution
  the force acting when the electron passes through the line joining two ions is in direction $\vec{\mu}\times\vec{v}$. 
  Averaging over the unit cell we showed in Ref. \cite{fm} that the net force is proportional to the averaged second derivative of the lattice potential and in direction
  $\vec{\mu}\times\vec{v}$, corresponding to an 'effective' magnetic field in direction $parallel$ to $\vec{\mu}$.

    \begin{figure}
\resizebox{8.0cm}{!}{\includegraphics[width=7cm]{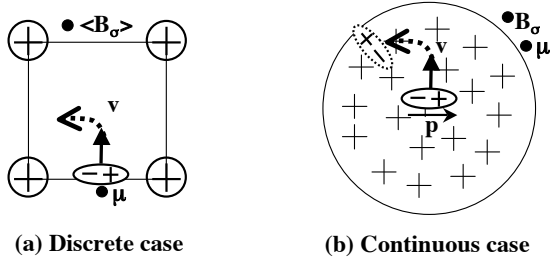}}
\caption{(a) The equivalent electric dipole for an electron with magnetic moment pointing out of the paper propagating upward
is attracted by the ion on the left and repelled by the ion on the right, resulting in a force pointing left. (b) The torque from the radially outward pointing electric field of the ions
changes the angular momentum of the electron so that it deflects to the left, aquiring angular momentum pointing out of the paper, i.e. parallel to $\vec{\mu}$. Similarly
an electron with magnetic moment pointing into the paper deflects in the opposite direction.}
\label{atom5}
\end{figure}
 
For the 'mesoscopic' scale under consideration ($2\lambda_L$)   it is more appropriate to consider a $uniform$ positive charge density, as shown
in Fig. 2(b). The electric field points radially outward,
\beq
\vec{E}=2\pi\rho\vec{r}
\eeq
with $\rho$ the ionic charge density. The torque by the electric field acting on the equivalent electric dipole changes the angular momentum
\bmath
\beq
\frac{d\vec{L}}{dt}=m_e\frac{d}{dt}(\vec{r}\times{\vec{v}})=\vec{\tau}=(\frac{\vec{v}}{c}\times\mu)\times\vec{E} 
\eeq
hence
\beq
 m_e\vec{r}\times\frac{d\vec{v}}{dt}=-2\pi \rho\vec{r}\times (\frac{\vec{v}}{c}\times\vec{\mu}) 
\eeq
\emath
which is equivalent to the effect of an 'effective' magnetic field in direction $parallel$ to $\vec{\mu}$
\beq
\vec{B}_\sigma=-2\pi\frac{\rho}{e}\vec{\mu}
\eeq
on the electron, also causing it to deflect in direction  $\vec{\mu}\times\vec{v}$, as shown in Fig. 2(b). Eq. (105) is the same as Eq. (11), since the superfluid
of negative charge density $en_s$ should respond to the positive ionic charge density $\rho=-en_s$.

The corresponding `unscreened' vector potential, defined by $\vec{\nabla}\times \vec{A}_\sigma=\vec{B}_\sigma$ is given by
\beq
\vec{A}_\sigma=\frac{\vec{B}_\sigma \times\vec{r}}{2}=-\pi\frac{\rho}{e}\vec{\mu}\times\vec{r}
\eeq
which yields
\beq
\vec{A}_\sigma=-\frac{\hbar}{4m_e c} \vec{\sigma}\times \vec{E} .
\eeq
for $\vec{E}$ given by Eq. (103).
The spin-orbit interaction term in the non-relativistic limit of the Dirac equation for an electron in an electric field $\vec{E}$ is
\bmath
\beq
H_{s.o.}=-\frac{e\hbar}{4m_e^2 c^2}\vec{\sigma}\cdot(\vec{E}\times\vec{p})=\frac{e}{m_e c}\vec{p}\cdot\vec{A}_\sigma
\eeq
where $\vec{p}$ is the momentum operator and $\vec{A}_\sigma$ is given by Eq. (107). Thus, we can understand the relation Eq. (18) between velocity and spin-orbit vector
potential as arising from minimization of the energy-momentum relation that arises from the non-relativistic limit of the Dirac equation
\beq
E(\vec{p})=\frac{p^2}{2m_e}+\frac{e}{m_e c}\vec{p}\cdot\vec{A}_\sigma
\eeq
\emath
with $\vec{p}=m_e\vec{v}_\sigma$ and $\vec{A}_\sigma$  given by Eq. (107).

It should be mentioned however that our conclusion is opposite to that of Chudnovsky\cite{chud} concerning the sign of the spin-orbit force acting on an electron moving in a positive background. 
Ref. \cite{chud} defines a spin-orbit vector potential as Eq. (107) {\it with opposite sign}, and concludes that the force is in opposite direction to that depicted in Fig. 2. 
We argue that it is physically clear that
the electron will deflect as shown in  Fig. 2, so that the electron orbital angular momentum will point antiparallel to its
spin angular momentum to minimize the spin-orbit energy, just like in atoms.  The difference between our conclusion and that of ref. \cite{chud} 
can be accounted for by the momentum of the electromagnetic field, which was calculated by Aharonov et al as \cite{comment}
\beq
P_{field}=\frac{1}{4\pi c} \int \vec{E}\times\vec{B}=\frac{1}{c}\vec{E}\times\vec{\mu}
\eeq
and  points in opposite direction to the mechanical momentum of the electron  in Fig. (2) and is twice as large. 

As electrons of opposite spin move radially outward a distance $2\lambda_L$ they acquire azimuthal velocity
\beq
\vec{v}_{\sigma}^0  =-\frac{e}{m_e c}\vec{A}_\sigma(r=2\lambda_L)=-\frac{\hbar}{4 m_e \lambda_L}\vec{\sigma}\times \hat{r} 
 \tag{13}
\eeq
where we have used $\rho=|e|n_s$\cite{sm} and Eq. (7).
 
 \section{Physical interpretation}

There are three key elements underlying the physical picture considered here that relate to the theory of hole superconductivity\cite{hsc}: (i) The wavefunction in the
 normal state has to be confined to short distances, which requires an almost full band (hole transport in the normal state);
 (ii) The carriers forming the superfluid are long wavelength $electrons$\cite{eh2}, with $negative$ $charge$\cite{eha}; (iii) In quantum mechanics, expansion of the
 wave function is associated with lowering of $kinetic$ $energy$, which is the driving force for hole superconductivity\cite{kinetic}.
 
 It is   interesting to note that an early explanation of superconductivity proposed by J.C. Slater in 1937\cite{slater} had some key elements in common
 with what is being proposed here.
 Slater proposed a model where ``the wave functions correspond to electrons which can wander for some distance through the metal'', and showed that
 {\it ``to produce superconductivity the orbits must be of order of magnitude of 137 atomic diameters''}.
 
 Consider an external magnetic field $\vec{H}$ applied to an electron orbiting with radius $r$ in a plane perpendicular to $\vec{H}$. Classically,
 in changing the field from $H$ to $H+dH$ the electron changes its speed by $dv=(e/m_ec)(rdH/2)$
 due to Faraday's law, and its orbital
 magnetic moment by $d\mu=(er/2c)dv$ antiparallell to the field, resulting in an energy increase $dE=Hd \mu$. When the field is increased from $0$ to $H$, the total
  energy increase is
 \beq
 \Delta E=\int _0^H dH' \frac{e^2}{4m_e c}r^2 H' =\frac{e^2}{8 m_e c^2} r^2 H^2
 \eeq
 The same result is obtained quantum-mechanically, with $r^2$ replaced by the mean square radius in the plane perpendicular
 to $\vec{H}$, $r^2\rightarrow <x^2+y^2>$. If there are $n_s$ electrons per unit volume in such orbits, the magnetic 
 susceptibility per unit volume is
 \beq
 \chi=-n_s\frac{\partial^2\Delta E}{\partial H^2}=-\frac{n_s e^2}{4 m_e c^2} r^2
 \eeq
 and for orbits of radius $r=2\lambda_L$, as proposed here,
 \beq
 \chi=-\frac{n_s e^2}{m_e c^2} \lambda_L^2=-\frac{1}{4\pi}
 \eeq
 using Eq. (7) for the London penetration depth. Eq. (112) is the condition for perfect diamagnetism. Instead, the Landau diamagnetic susceptibility
 in the normal state is given by Eq. (111) with $r=k_F^{-1}$, with $k_F$ the Fermi wavevector.

 Similarly, Eq. (110) yields that the increase in energy per unit volume for $n_s$ electrons per unit volume in orbits of
 radius  $r=2\lambda_L$  is
 \beq
 u\equiv n_s\Delta E=\frac{n_se^2}{8 m_e c^2} (2\lambda_L)^2 H^2=\frac{H^2}{8\pi}
 \eeq
 again using Eq. (7), 
 so that the system will remain in this state as long as the 'condensation energy' density of the state is greater than the energy cost
 $H^2/8\pi$. This is of course the condition that defines the thermodynamic critical field $H_c$\cite{tinkham}.

   Slater's arguments\cite{slater} proceeded similarly. He pointed out that the radius of the orbit in the susceptibility expression
  Eq. (111) should be such that $\chi=-1/4\pi$ for perfect
  diamagnetism; taking $n_s=1/d^3$, with $d^3$ the volume per
 superconducting electron, and setting $d=2a_0$, with $a_0=\hbar^2/m_e e^2$ the Bohr radius, yields
 \beq
 \frac{r^2}{d^2}=\frac{m_e c^2}{e^2}\frac{d}{\pi}=\frac{2}{\pi}(\frac{\hbar c}{e^2})^2        
 \eeq
 hence Slater concluded that ``the orbits must be of order of magnitude of 137 $(=\hbar c/e^2)$ atomic diameters''.
 
 Furthermore, Slater proposed as a criterion for the critical magnetic field that will destroy superconductivity to
 compare the Landau level spacing of energy levels for fixed z-component of momentum in an external field $H$ 
 \beq
 \Delta E_{ll}=\frac{e\hbar}{m_e c}H
 \eeq
 to the spacing of discrete energy levels in a 'box' of size 137 atomic units. Instead, we argued in ref.\cite{sm} that
 superconductivity will be destroyed when the external magnetic field completely stops the spin current orbital motion of the
 electron with spin antiparallell to $H$, i.e. for
 \beq
 H=B_\sigma
 \eeq
 with $B_\sigma$ given by Eq. (15a). The Landau level energy spacing for such a magnetic field is
 \beq
  \Delta E_{ll}=\hbar \omega_c =\frac{e\hbar}{m_e c}B_\sigma=\frac{\hbar^2}{4m_e \lambda_L^2}
  \eeq
  which is of the same order of magnitude as the energy level spacing of electrons confined to a region of size
  $\sim 2\lambda_L$. Thus our criterion is essentially equivalent to Slater's criterion. An equivalent form of this criterion is
  that superconductivity will be destroyed when the energy increase for electrons in  a magnetic field $H$ for orbits
  of radius  $r=2\lambda_L$, Eq. (110), becomes comparable to the spacing of energy levels in a box of such size.

 Of course Slater's scenario left several questions unanswered that our present scenario addresses, namely: 
 (i) what happens to those orbits when there is no external magnetic field applied?
 As we showed in \cite{sm}, orbits with radius $2\lambda_L$  arise from the spin-orbit interaction under the
 constraint that the orbit should enclose precisely one flux quantum of spin orbit flux, so that the wave function is single-valued. And (ii) how do the orbits arise
 when a metal is cooled into the superconducting state, and how is the Meissner current generated?
 It is the process of  $expansion$ of the orbits from negligible radius (of order of a lattice spacing) to radius $2\lambda_L$ that
 generates $dynamically$ the azimuthal velocities required. 
  
 It is also interesting to note that the physics of superconductivity discussed here provides a natural 'classical' explanation for the existence of 
 macroscopic phase coherence in superconductors. The fact that electrons traverse overlapping orbits of radius $2\lambda_L$ implies that the orbits of many electrons cross 
 the orbit of any given electron, so the motions needs to be synchronized: to avoid collisions the angular position of the i-th electron $\theta_i(t)=\omega_c t+\theta_{i0}$
 has to bear a definite relation with the angular positions of the electrons in the overlapping orbits, that persists over time.   It is also not possible to change one orbit without affecting 
 the synchronization with all the others, which gives an intuitive understanding to the notion of 
  'rigidity' of the superconducting state\cite{london}.  Finally, it is interesting to note that several workers in the pre-BCS era proposed mechanisms of superconductivity based on the notion of
  'spontaneous currents'  that would exist in the superconductor in the absence of applied external fields\cite{spontaneous}.
  However these were charge currents rather than spin currents.
 
 \section{Angular momentum}
 Within our model each electron in the superfluid moves in a circular orbit of radius $2\lambda_L$, with electron spin perpendicular to the plane of the orbit in
 direction antiparallel to the orbital angular momentum, and velocity given by Eq. (13). Hence the mechanical orbital angular momentum of the  i-th electron  is
 \beq
 \vec{l}_{i,orb}=m_e(2\lambda_L)\hat{r}\times\vec{v}_\sigma= -\frac{\hbar}{2}\vec{\sigma}
 \eeq
 where the last equality follows from Eq. (13). The electron spin angular momentum is
 \beq
 l_{i,spin}=+\frac{\hbar}{2}\vec{\sigma}
 \eeq
and consequently its total angular momentum (orbital plus spin) is {\it exactly zero}. Thus, the transition to superconductivity can be thought of as a ``quenching'' of the
 electron's spin angular momentum by the development of an opposite-pointing orbital angular momentum of the same magnitude.
 
Let us consider the total mechanical angular momentum for electrons of spin $\sigma$ in a superconducting cylinder of radius $R$ and height $h$, choosing as usual
 the spin quantization axis parallel to the cylinder axis. The total number of electrons with spin $\sigma$ for superfluid density $n_s$  in the entire volume
 $V=\pi R^2 h$   is
 \beq
 N_{vol,\sigma}=\frac{n_s}{2}V
 \eeq
Hence the total mechanical angular momentum carried by electrons with spin $\sigma$ is the product of Eq. (120) and (118):
 \beq
 \vec{L}_{vol,\sigma}=N_{vol,\sigma}\vec{l}_{i,orb}=-n_s V m_e \lambda_L v_\sigma \vec{\sigma}=-\frac{n_s}{2} V \frac{\hbar}{2}\vec{\sigma}
 \eeq
 On the other hand, the orbital angular momentum for an electron of spin $\sigma$ in the surface layer of thickness $\lambda_L$ is
 \beq
 \vec{l}_{surf}=m_e R \hat{r}\times\vec{v}_\sigma=\frac{R}{2\lambda_L} \vec{l}_{i,orb}
  \eeq
 and the total number of electrons of spin $\sigma$ in the surface layer of thickness $\lambda_L$ is
 \beq
 N_{surf,\sigma}=2\pi R \lambda_L h \frac{n_s}{2}=\frac{2\lambda_L}{R}V\frac{n_s}{2}=\frac{2\lambda_L}{R}N_{vol,\sigma}
 \eeq
 so that the total mechanical angular momentum carried by the electrons of spin $\sigma$ in the surface layer is
 \beq
 \vec{L}_{surf,\sigma}=N_{surf,\sigma}\vec{l}_{surf}=N_{vol,\sigma}\vec{l}_{i,orb}=\vec{L}_{vol,\sigma},
 \eeq
 the same as Eq. (121). So we may think of the total mechanical angular momentum in two equivalent ways: (i) each superfluid electron in the bulk carries mechanical angular momentum
 $\hbar/2$, or (ii) the superfluid electrons in the surface layer of thickness $\lambda_L$ each carry a mechanical angular momentum 
 $R\hbar/4\lambda_L$ (resulting from their orbital motion with speed $v_\sigma$ and radius $R$), and the electrons in the interior carry no mechanical 
 angular momentum. The total mechanical angular momentum of the superfluid electrons of each spin component in the surface layer equals in magnitude  the 
 total $spin$ angular momentum of the superfluid electrons of that spin component in the entire volume.

 If we cool the system into the superconducting state in the presence of an applied magnetic field $\vec{B}$, the electrons of both spin orientations acquire
 an extra contribution to their orbital angular momentum as they expand to their orbits of radius $2\lambda_L$. 
 The azimuthal velocity acquired is
 \beq
 \vec{v}_\phi=\frac{e\lambda_L}{m_e c}\hat{r}\times\vec{B}
 \eeq
 due to the action of the Lorentz force as the orbits expand to radius $2\lambda_L$\cite{sm}. $\vec{v}_\phi$ increases the speed of the electrons with $\vec{\mu}$ parallel to $\vec{B}$, and decreases the speed of
 the electrons with $\vec{\mu}$ antiparallel to $\vec{B}$. The extra angular momentum per electron
 due to the magnetic field  is
 \beq
 \vec{l}_B=m_e (2\lambda_L)\hat{r}\times\vec{v}_\phi=-\frac{2e \lambda_L^2}{c}\vec{B}
 \eeq
 and the total extra angular momentum acquired by the electrons is obtained by multiplying $\vec{l}_B$ by the total number of electrons:
 \beq
 \vec{L}_e=-\frac{2en_s V \lambda_L^2}{c}\vec{B}=-\frac{m_e c R^2 h}{2e}\vec{B}
 \eeq
 as expected, since
 \beq
 \vec{L}_e=\frac{2m_e c}{e}\vec{m}
 \eeq
 for electrons, with $\vec{m}=\pi R^2 h \vec{M}$ the induced magnetic moment and $\vec{M}=-\vec{B}/4\pi$  the required magnetization to cancel the magnetic field
 in the interior. Hence, just like for the angular momentum in the spin current, the angular momentum in the Meissner current in the surface layer 
 can be interpreted as arising from the orbits of radius $2\lambda_L$ of
 each electron in the bulk.
 
  \begin{figure}
\resizebox{8.0cm}{!}{\includegraphics[width=7cm]{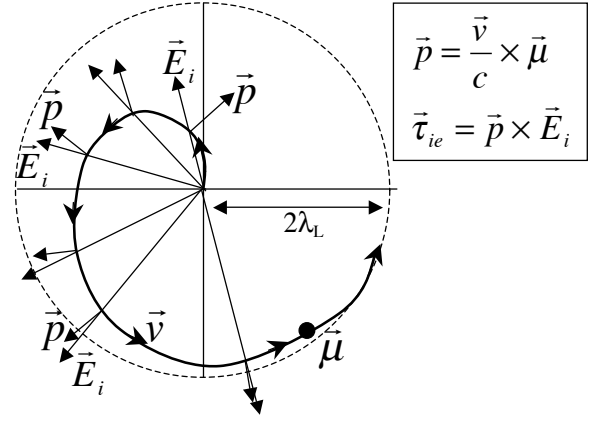}}
\caption{The thick line shows the trajectory of an electron of magnetic moment $\vec{\mu}$ pointing out of the paper starting at the origin till it reaches 
the circumference of radius $2\lambda_L$. The equivalent electric dipole moment $\vec{p}$ points perpendicular to the trajectory and outward, 
the ionic electric field $\vec{E}_i$ points radially outward and the cross product $\vec{p}\times\vec{E}_i$ is parallel to $\vec{\mu}$.   
The torque exerted by the lattice on this electron $\vec{\tau}_{ie}=\vec{p}\times\vec{E}_i$ points out of the paper and goes to zero as the electron reaches the circular trajectory.
For the electron of opposite spin, the torque points into the paper and the trajectory is a mirror image across the vertical direction. In the presence of a magnetic
field pointing out of the paper the torque on this electron becomes larger and the one on the electron of opposite spin becomes smaller. Hence there is a net torque exerted by the
ions on the electrons pointing out of the paper, and a reaction torque exerted by the electrons on the ions pointing into the paper that causes the body as a whole 
to rotate clockwise. }
\label{atom5}
\end{figure}
 
 How is this electronic angular momentum compensated? We raised this puzzle in ref. \cite{lenz}: in the conventional theory of superconductivity the charge 
 distribution is homogeneous, so there is no angular momentum in the electromagnetic field. The electrons in the Meissner current carry mechanical angular
 momentum $\vec{L}_e$, but there is no mechanism in the conventional theory to generate a compensating angular momentum of the ions.
 
This conundrum is naturally resolved in the present scenario. Recall that in the absence of magnetic field the $2\lambda_L$ circular orbits arise from the
 interaction of the  magnetic moment moving outward 
 with the positive ionic lattice. The physics is illustrated schematically in Fig. 3. The ions exert a torque on the electron 
 \beq
 \vec{\tau}_{ie}=\vec{p}\times\vec{E}_i
 \eeq
 where $\vec{E}_i$ is the ionic electric field and $\vec{p}$ is the equivalent electric dipole  Eq. (102). 
 As the ions exert the torque on the electrons the ionic lattice is subject, by Newton's third law, to an equal and opposite torque:
 \beq
 \vec{\tau}_{ei}=-\vec{\tau}_{ie}
 \eeq
 In the absence of applied magnetic field, $\vec{\tau}_{ei}$ is equal and opposite from electrons with spin up and spin down, hence there is no 
 net torque on the ions. However, in the presence of $\vec{B}$ the velocity of both spin electrons is modified according to Eq. (125),
 and hence the net $\vec{\tau}_{ei}$ is no longer zero: as the electrons acquire the extra angular momentum Eq. (126), the extra torque on the ions
 points opposite to the direction of $\vec{B}$ and generates an equal and opposite angular momentum for the lattice, $\vec{l}_{ions}=-\vec{l}_B$ per 
 electron, and a total ionic angular momentum
 \beq
 \vec{L}_i=-\vec{L}_e
 \eeq
 so it would appear that the total angular momentum is conserved.

 However that is not quite the whole story. Because the resulting charge distribution is inhomogeneous, there will also be some angular momentum stored in the
 electromagnetic field:
 \beq
 \vec{L}_{field}=\frac{1}{4\pi c}\int d^3r \vec{r}\times(\vec{E}\times\vec{B})
 \eeq
which  we can estimate as
  \beq
 \vec{L}_{field}=-\frac{V\lambda_L}{2\pi c} E_m \vec{B}
 \eeq
 since the region where both electric and magnetic fields are nonzero is only the surface layer of thickness $\sim\lambda_L$.
Using Eqs. (130), (124), (33) and (7)
 \beq
 L_{field}=\frac{\hbar}{8 m_e c \lambda_L}L_e=\frac{1}{8\pi} (\frac{\lambda_c}{\lambda_L}) L_e
 \sim 10^{-6}L_e
 \eeq
 with $\lambda_c=h/m_e c$ the Compton wavelength. Hence we conclude that the spin-orbit interaction can account for $99.9999\%$ of the angular momentum 
 conservation puzzle\cite{lenz} through Eq. (131), but we are still missing $0.0001\%$!
 
This 'missing' angular momentum is of course the    tiny bit of extra electronic angular momentum that is acquired by the electrons in $\rho_-$ that moved $outward$ cutting magnetic
 field lines near the surface in the process. The change in azimuthal velocity of an electron near the surface moving outward a distance $\lambda_L$ in a magnetic field $B$ due to the Lorentz force is
 \beq
 \Delta v=\frac{e}{m_ec} \Delta A= 
 \frac{e}{m_ec}\frac{\Delta \phi}{2\pi R}=\frac{e\lambda_L B}{m_ec}
 \eeq
 and the acquired angular momentum is 
 \beq
 \vec{l}=-\frac{e}{c}R\lambda_L\vec{B}
 \eeq
 parallel to the magnetic field. The number of electrons that moved out to the region within $\lambda_L$ of the surface and in the process cut through magnetic field lines can be estimated as
 \beq
 N_{e,out}=\frac{\rho_-}{e}2\pi R \lambda_L h
 \eeq
 hence the mechanical angular momentum gained by the outflowing $\rho_-$ charge  is
 \beq
 \vec{L}'_e=N_{e,out} \vec{l}=-\frac{2\rho_-V\lambda_L^2}{c}\vec{B}=\frac{V\lambda_L}{2\pi c} E_m \vec{B}
 \eeq
 which is equal and opposite to the angular momentum of the field Eq. (133). (Note that Eq. (2a) was used in deriving Eq. (138)). 
Consequently, angular momentum is conserved when a metal is cooled into the superconducting state in the presence of a magnetic field, 
since the total angular momentum above $T_c$ is zero, and from Eqs. (131), (133) and (138)
 \beq
 \vec{L}_e+\vec{L}'_e+\vec{L}_{field}+\vec{L}_i=0   .
 \eeq

 \section{Type I versus type II materials}
 
 In a type II superconductor, magnetic flux penetrates the body for $H_c>H_{c1}$ and divides itself into filaments,
 each carrying a flux quantum $\phi_0$. In the vortex core of size $\xi$, the 'coherence length', the system is normal.
 Instead, in a type I superconductor the flux is excluded until the applied field reaches the value $H=H_c$, the 
 thermodynamic critical field, and for $H>H_c$ the entire material becomes normal. The thermodynamic critical
 field is given by
 \beq
 H_c=\sqrt{\frac{3}{2}} \frac{\hbar c}{\pi \xi_0\lambda_L}\sim H_{c1} \frac{\lambda_L}{\xi_0}
 \eeq
 with $\xi_0$ the Pippard-BCS coherence length\cite{tinkham}.
 A similar relation holds at finite temperatures between the Ginzburg Landau coherence length $\xi$ and the penetration depth.
 We will not distinguish here between $\xi$ and $\xi_0$. For type II superconductors, $\xi<\lambda_L$, hence $H_{c1}<H_c$.
 
 We propose the following interpretation of type I versus type II behavior, depicted in Fig. 4. Up and down spin electrons have orbits of radius
 $2\lambda_L$, but the orbits coincide only in the extreme type II limit where $\lambda_L>>\xi$. {\it $\xi$ represents the (average)
 distance between the centers of the orbits of spin up and spin down components of the Cooper pair}.

   \begin{figure}
\resizebox{8.0cm}{!}{\includegraphics[width=7cm]{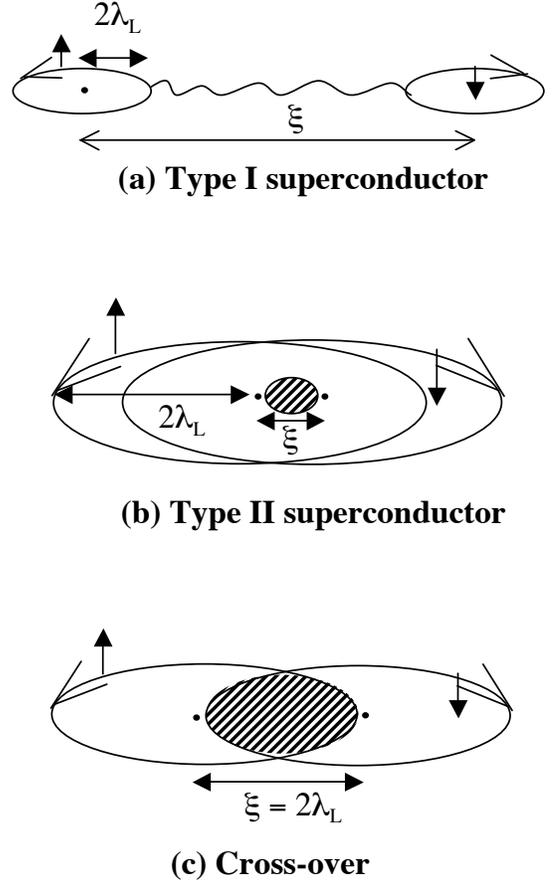}}
\caption{Type I versus type II materials. $\xi$ is the distance between the centers of the orbits of magnetic moment up electron (left circle)
and magnetic moment down electron (right circle). The small dots denote the centers of the orbits, the up and down arrows the
direction of the electron's magnetic moment, their direction of motion is indicated with a horizontal arrow.
In type I materials ((a)), $\xi>2\lambda_L$ and  the orbits don't overlap. 
In type II materials ((b)) with $\xi>2\lambda_L$, a normal vortex core can be enclosed by both orbits.
 (c) denotes the cross-over situation, with $\xi=2\lambda_L$ }
\label{atom5}
\end{figure}

 In a type I superconductor (Fig. 4(a)), $\xi>>\lambda_L$ and the orbits of $\uparrow$ and $\downarrow$ electrons are
 disjoint. Because orbits are time-reversed partners, a magnetic field cannot thread one of the orbits and not the other.
 Instead, in a type II material with $\xi<<\lambda_L$ (Fig. 4(b)) magnetic flux can
 be enclosed by both orbits simultaneously. When the magnetic field first enters a type II superconductor, each orbit will enclose a single flux quantum $\phi_0$
 and the magnetic field at the center of the vortex is $H_{c1}$. In the mixed state, at the core of radius $\xi=$ distance between the centers of the
 orbits, the magnetic field can have any value between $H_{c1}$ and $H_{c2}$ and there will be several vortex cores within the orbits of a Cooper pair with  the total flux threaded by each orbit an integer
 multiple of $\phi_0$. 
 The cross-over between type I and type II  regimes occurs for
 $\xi=2\lambda_L$ (Fig. (4c)), where the overlapping part of the orbits can just enclose the vortex core. 
 When the centers of the orbits are at distance $\xi>2\lambda_L$ (type I), the
 magnetic flux will destroy the Cooper pairs altogether, but depending on the sample shape there can be intertwined regions of normal and superconducting phases (intermediate state).
 
 Fig. 5 shows schematically type I and type II superconductors in a magnetic field. In type I superconductors with non-zero demagnetizing factor there will be 
 normal regions usually of laminar shape intertwined with the superconducting regions. The uncompensated orbits in the neighborhood of the normal regions and
 vortices will give rise to
 spin current and excess negative charge  similar to the behavior near the surface.

      \begin{figure}
\resizebox{9.0cm}{!}{\includegraphics[width=7cm]{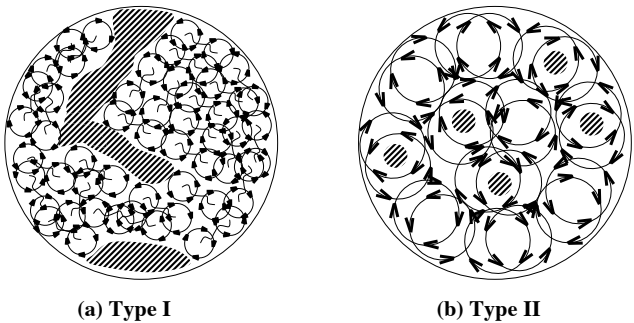}}
\caption{Type I versus type II materials in a magnetic field (schematic).  (a) The wavy lines connect the centers of the two members of a
Cooper pair. In the intermediate state, normal regions exist with laminar or other shapes (shaded regions) of size larger than $\xi$ where Cooper pairs
are destroyed and the magnetic field is $H_c$. (b) $\xi<<\lambda_L$ and the orbits of the two members of a Cooper pair almost overlap. The vortex cores (shaded regions) are within the
orbits of the two members of a Cooper pair and have magnetic field $H_{c1}$ when vortices don't overlap.  }
\label{atom5}
\end{figure}

 This interpretation also provides a rationale for the form of the critical field Eq. (140). For $\xi>\lambda_L$, the 'confinement'
 region for an electron in the Cooper pair is $\xi$ rather than $2\lambda_L$. Following Slater's reasoning\cite{slater},
 we equate the diamagnetic energy cost for the electron in an orbit of radius $2\lambda_L$ in the presence of an external
 field $H$ to the spacing of energy levels in a box of length $\xi$:
 \beq
 \Delta E_H=\frac{e^2}{8m_e c^2}(2\lambda_L)^2 H^2\sim\frac{\hbar^2}{2m_e\xi^2}
 \eeq
 and obtain
 \beq
 H\sim\frac{\hbar c}{e\lambda_L \xi}
 \eeq
 which is essentially the thermodynamic critical field Eq. (140). Thus, in a type I material superconductivity is not destroyed
 when the magnetic field stops the spin current\cite{sm} (since $H_{c1}>H_c$ here) but rather when the
 diamagnetic energy cost is large enough that the perturbation significantly mixes the unperturbed energy levels,
 i.e. disturbs the wave function so that it is no longer 'rigid'.

 What is the expelled charge density $\rho_-$ in type I materials? The conclusion that $E_m$ is given by Eq. (33) ($E_m\sim H_{c1}$)  is not valid
 because the electrodynamic response is non-local. Furthermore, the electrostatic energy cost Eq. (44) would be much
 larger than the condensation energy in this case. The amount of $\rho_-$ 'needed' to sustain the Meissner current up to
 $H=H_c$ is, according to Eq. (41), simply $\rho_-=-H_c/4\pi$, so we conclude that for type I materials
 \bmath
 \beq
 E_m=H_c
 \eeq
 \beq
 \rho_-=-\frac{H_c}{4\pi \lambda_L }=\sqrt{\frac{3}{2}}\frac{1}{4\pi^2} \frac{\hbar c}{e\lambda_L^2\xi_0}
 \eeq
 \emath
 rather than Eqs. (33) and (34) applicable to type II materials.
 Because of the non-locality we expect the spin current density near the surface in type I materials will no longer be given by Eq. (16), but rather
 \beq
 \vec{J}_\sigma=-\frac{\rho_- c}{2}\vec{\sigma}\times\hat{r}
 \eeq
 In type I materials Eq. (144) is smaller than Eq. (16) by a factor $\lambda_L/\xi$.

Next  we address the fact that disorder is known to increase the value of $\lambda_L$ and turn a 
 type I into a type II material. This can be easily understood in the picture discussed here.
 In the presence of defects that weaken or destroy superconductivity locally,
 the superconductor will expel negative charge towards those regions, thus depleting the magnitude of $\rho_-$ near the
 outer surface and decreasing the ability of the superconductor to shield external magnetic fields. These internal weak regions (grain boundaries, defects, etc) will have excess negative charge
 $and$ accompanying spin current orbiting around them, as shown schematically in Fig. 6. If we cool this
 superconductor in the presence of a pre-existent magnetic field, it will be unable to expel the magnetic field from
 those regions, thus resulting in an incomplete Meissner effect.

   \begin{figure}
\resizebox{8.0cm}{!}{\includegraphics[width=7cm]{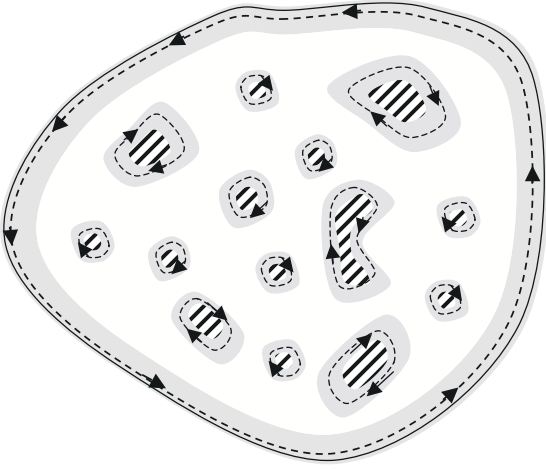}}
\caption{Schematic depiction of a superconductor with strong disorder in the absence of applied magnetic field. Defects, grain boundaries, vacancies, etc. will result
in patches of normal regions (hatched areas) surrounded by spin currents (dashed lines, with arrows pointing in the direction of flow of electrons with
magnetic moment pointing $out$ of the paper) and excess negative charge density (gray areas). 
The figure also shows the excess negative charge and spin current near the surface and the spill-over beyond the surface (denoted by the full line). 
The equivalent electric dipoles (Eq. (102)) point outward near the outer surface and
towards the normal regions in the interior.
If the system is cooled in the presence of a magnetic field, magnetic flux will be trapped in the hatched regions and a charge current will flow around
those regions together
with the depicted  spin currents. The smallest normal regions have diameter of a coherence length and enclose one flux quantum of spin-orbit flux in
the absence of applied magnetic field. .}
\label{atom5}
\end{figure}

 This picture also suggests that in extreme type II materials there will be an inhomogeneous charge distribution and
 spin current distribution arising from inhomogeneities and disorder also in the absence of applied magnetic field.
 Such inhomogeneities are observed in the underdoped regime of high $T_c$ materials\cite{davis}, which are expected  to be in the extreme type II limit
 (largest $\lambda_L$, smallest $\xi$)\cite{holesc}. The   existence of spin currents associated with regions of charge inhomogeneity predicted by our model has not yet been
 experimentally tested.
 
Finally, recall that we argued in Sect. V that the equations for the spin current spatial dependence should include the constant term $\vec{J}_{\sigma_0}$, giving
 rise to  a singularity in the deep interior.  Presumably this implies that even in the absence of disorder
at least one  'vortex', i.e. a normal region of size $\xi$ surrounded by spin current, has to exist in the deep interior of any type II  superconductor. 
This would imply  that any type II  superconductor is topologically  a torus of genus larger or equal to 1 even
in the absence of applied magnetic field, hence  that a Meissner effect with $100\%$ flux expulsion can never be attained
in a type II superconductor no matter how small the applied external field: at the minimum one trapped magnetic flux quantum
will always remain in the interior.

\section{Superconducting material parameters}

\begin{table*}
\caption{ 
  Properties of superconducting elements and compounds. The superfluid carrier density $n_s$ is extracted from the measured penetration
depth $\lambda_L$ using expression Eq. (7), so it corresponds to what is usually   interpreted as $n_s m_e/m^*$, with $m^*$ the effective mass.
 The spin current density $J_{spin}$ is defined by Eq. (144) and includes the contribution from both spin components.
 }
\begin{tabular}{l || c | c  | c | c  | c | c |c  | c | c | c | c}
Material & $T_c (K)$& $\lambda_L (\AA)$ & $H_c (G)$ &$H_{c1} (G)$  &  $n_s (e/\AA ^3) $ & $n_s (e/i) $  &$10^6\rho_-/en_s$  & $E_m(V/cm)$  & $v_{\sigma}(cm/s)$  & 
$J_{spin} (10^6 A/cm^2)$ \cr
\tableline
Cd & 0.56 & 1300  &  30   &97    &0.00167&0.0361&0.23 & 9,000 & 22,275 & 1.83   \cr
Zn & 0.875 & 290  &  53  & 1955    &0.0336 & 0.510 &0.090  &15,900  &99,855  &14.6\cr
Al & 1.14 & 500 &  105  & 658  & 0.0113  &   0.188   &0.31  &31,500  & 57,916 &16.7     \cr
In &3.40  & 640  &  293   & 401    &0.0069&0.180  & 1.10 &87,900  &45,247  & 36.4    \cr
Sn & 3.72 & 510  &  309   &632   &0.0109  & 0.293 & 0.92  & 92,700 &56,780  & 48.0    \cr
Hg & 4.15 & 410  &  412    &978   &0.0168 &0.412  &0.99 & 123,600 & 70,629 &   80.2  \cr
Pb & 7.19 & 390  &  803   & 1080   &0.0186 & 0.561 &1.83 &240,900  &74,251  &  163.7   \cr
Nb & 9.50 & 400  &  1980   &1028    &0.0176 &0.324  &2.41 &308,400  & 72,395 &  205  \cr
$MgB_2$   & 39.2   &   1800  & 2600 & 51   &0.00087  &0.0084  & 0.54  &15,300 & 16,088 & 2.26   \cr
$La_{1.85}Sr_{.15}Cu_2O_4$&37.3 & 2500 &2370   &26   &0.00045 & 0.0071 & 0.38  & 7,800 & 11,583 & 0.826    \cr
$YBa_2Cu_3O_7$ &  90    & 1500 &6450   &73   &0.0013 &0.017  &0.62& 21,900& 19,305 & 3.88    \cr
\tableline
 \end{tabular}

\end{table*}

From experimentally measured values of the penetration depth and critical fields we can infer 
the magnitude of the maximum internal electric field $E_m$, surface charge density $\rho_-$,
velocity of the spin current carriers at the surface, and the spin current density flowing near the surface.
We use the expressions:
\beq
H_{c1}=\frac{\hbar c}{4 e \lambda_L^2}=\frac{1.64\times 10^8}{\lambda_L(\AA)^2} Gauss
\eeq
\bmath
\beq
E_m=H_{c1}(G)\times 300 V/cm
\eeq
for type II materials ($H_{c1}<H_c$) and 
\beq
E_m=H_{c}(G)\times 300 V/cm
\eeq
\emath
for type I materials.
 \bmath
\beq
n_s=\frac{m_e c^2}{4\pi e^2 \lambda_L^2}=\frac{2824}{\lambda_L(\AA)^2}\frac{electrons}{\AA^3}
\eeq
or
\beq
n_s( electrons/ion ) = n_s \frac{\bar{A}} {0.602\rho (gr/cm^3)}
\eeq
 \emath
with $\bar{A}$ the average mass number of the compound,
\beq
\frac{\rho_-}{e}=\frac{E_m}{4\pi \lambda_L e}=5.524\times 10^{-11}\frac{E_m(V/cm)}{\lambda_L(\AA)} \frac{electrons}{\AA ^3}
\eeq

\beq
v_\sigma=\frac{\hbar}{4 m_e \lambda_L}=\frac{2.896\times 10^7}{\lambda_L(\AA)}cm/s
\eeq
\beqn
J_{spin}&=&J_{\sigma=+1}-J_{\sigma=-1}=\rho_- c  \nonumber \\
&=&\frac{\rho_-}{e}(\frac{el}{\AA ^3})\times 4.802 \times 10^{15} Amps/cm^2
\eeqn

We list the values for a variety of materials including type I and type II superconductors in Table I. From this table we learn that materials with the
largest values of $\rho_-$, $E_m$ and $J_{spin}$ are $Nb$, $Pb$ and $Hg$. They are close to the cross-over between type I and type II regimes
and have relatively high $T_c$. Hence these materials should be favored in experiments aimed at detecting this unconventional physics.
Instead, the higher $T_c$ materials known to date (last entries in Table I and similar) are strongly type II and the magnitude of the spin current and 
associated quantities is substantially smaller.

\section{Summary and discussion}

In this paper we have extended the electrodynamic equations for the charge sector of superconductors proposed in Ref.\cite{edyn} to describe the electrodynamics of the
spin sector, based on the proposal of  ref.\cite{sm} that a spontaneous macroscopic spin current flows within a London
penetration depth of the surface of superconductors, in the absence of applied external fields,  with carrier speed given by the universal form Eq. (13). 
We have shown here that the
space and time dependence of the spin current can be described within the same four-dimensional framework developed earlier to describe the
charge electrodynamics within the theory of hole superconductivity, which predicts that negative charge is expelled from the interior to the
surface when a metal goes superconducting.  The extension of the theory to describe the spin sector is necessitated by the fact that an interior electric field exists in 
superconductors within this theory. Instead, in the conventional London theory there is no electric field in the interior of superconductors, hence no separate
electrodynamic description of the spin sector is required.

The formalism led us   to the determination of the magnitude of the expelled negative charge, which had been left undetermined
in our earlier work. Remarkably, the maximum electric field in the superconductor resulting from the charge expulsion was found to have identical magnitude as the
spin-orbit field that gives rise to the spontaneous spin current, $and$ to have  similar magnitude as $H_{c1}$, the conventional lower critical field of 
type II superconductors.  For type I superconductors, we conjectured that
the maximum electric field in the interior is $H_c$.  In addition, we found a remarkable connection between the magnitude of excess charge at any point in space and time  and the speed of the superfluid at that point,
\beq
\rho_\sigma c=\frac{e n_s}{2} v_\sigma
\eeq
which says that $any$ superfluid current (charge or spin) can be understood as originating in the {\it excess charge} moving at the speed of light.
We also found a remarkable parallel between the kinetic energy of the carriers due to   charge and spin currents and the respective energy densities of
magnetic and electric fields near the surface:
\bmath
\beq
\frac{1}{2} m_e (v_s)^2 n_s=\frac{B^2}{8\pi}
\eeq
\beq
\frac{1}{2} m_e (v_\sigma^0)^2 n_s=\frac{E_m^2}{8\pi}
\eeq
\emath
All of this  hints at  deeper underlying physics   that remains to be uncovered.

The macroscopic spin current originates in the microscopic physics proposed in Ref. \cite{sm}: that superfluid electrons reside in circular orbits of radius
$2\lambda_L$, their velocity given by Eq. (13), hence up and down spin electrons orbit in opposite direction with orbital angular momentum antiparallel to their spin angular
momentum and of the same magnitude,   $\hbar/2$. The state arises from the interaction of the electron with the positive ionic background, i.e. a spin-orbit interaction, as the wavefunction
expands in the transition to the superconducting state to lower the kinetic energy of confinement of the antibonding electrons at the
top of a nearly filled band.  As shown in ref.\cite{sm}, the expansion from a small orbit
to a mesoscopic  orbit of radius $2\lambda_L$ provides a $dynamical$ explanation for the Meissner effect as well as for the dynamical origin of the spin current.

Furthermore, the total mechanical angular momentum carried by each component of the spin current in the surface layer was found to be identical to the aggregate sum of the $spin$ angular momenta of that spin component of the  superfluid electrons in the entire volume. This follows from  the fact that the orbital angular momentum for each electron in its orbit of radius $2\lambda_L$  is   $\hbar/2$. It implies that  
  the entire superconductor is a single giant Cooper pair, with its 
giant spin current components reflecting as well as quenching the aggregate sum of the superfluid electron  spins.

We also found that our equations predict that a significant amount of negative charge spills out of the superconductor. This effect was
anticipated in our earlier work, and we believe it may play an important role in the proximity effect. It also clearly illustrates the tendency of superconductors to 
get rid of their excess negative charge (antibonding electrons at the top of the Fermi distribution)  predicted by the theory of hole superconductivity, which is also reflected in the predicted tunneling asymmetry of universal sign\cite{tunneling} (larger tunneling current for $negatively$ biased superconductor).

The fact that  in the proposed scenario   spin-orbit coupling is an essential ingredient of superconductivity also provides a natural explanation for
the mechanism by which the ionic lattice
picks  up the `missing' mechanical angular momentum\cite{lenz} when  a metal is cooled into the superconducting state in the presence of a magnetic field.
Conventional BCS-London theory neither provides an explanation for how the electrons in the Meissner current acquire their mechanical angular momentum, nor for
how the ions acquire a compensating angular momentum in the opposite direction\cite{lenz}. 
In our earlier attempt to explain the angular momentum conservation puzzle\cite{lenz} we could not account for a complete Meissner effect precisely because the spin-orbit interaction
mechanism to transfer angular momentum to the lattice was not included. For a $99\%$ Meissner effect, we had to assume a charge expulsion at least $3$ orders of magnitude larger
than the values discussed here (Eq. (27) of ref.\cite{lenz}).

Finally, we discussed a new 'geometric' interpretation of type I and type II superconducting regimes and of  the effect of disorder in superconductors based on the present model,
and we proposed that simply connected type II superconductors cannot exist according to this theory.

Note however that we have assumed at various points in this paper that the size of the superconductor is much larger than $\lambda_L$ (e.g. Eqs. (2) and (3)).
For superconducting samples of dimensions comparable or smaller than the penetration depth we expect the fundamental equations of Sect. VII to remain valid,
but several features will change: in particular, the spin current at the surface will be smaller than the universal form Eq. (13), and the maximum electric field $E_m$ will be 
smaller than the value given by Eq. (33)\cite{prb03}. Note also that our treatment assumed local electrodynamics and non-local corrections will be important  in  strongly type I superconductors.

The hypothesis  that electrons in superconductors reside in orbits of diameter hundreds of lattice spacings was originally
proposed by Slater\cite{slater}. We have shown that the radius of these `Slater orbits' has to be precisely $2\lambda_L$ to give rise to perfect diamagnetism.
 Indeed, as argued by Slater, large orbits provide a simple and compelling explanation for why superconductors cannot tolerate the presence of a magnetic field in their interior: it simply costs
too much energy, proportional to the area subtended by  these mesoscopic orbits.
Type I superconductors have only two ways to deal with this: either expel the magnetic field from their interior, paying the associated electromagnetic energy price, or if the price is too high, become normal, collapsing the mesoscopic orbits to microscopic Landau orbits of radius $k_F^{-1}\sim$ lattice spacing and pay the associated kinetic energy price. 
Type II superconductors have a third way: enclose within the
mesoscopic electron orbits tubes of magnetic flux where the system is normal. Type II superconductors are able to do this because the two members of a Cooper pairs have 
the centers of their $2\lambda_L$ orbits
sufficiently close to each other that they can enclose within them the same flux tube.

 Experiments should be able to detect the existence of the predicted spin current near the surface of superconductors.
 We have given numerical estimates for the magnitude of the spin current and other associated quantities for 'conventional' and other superconductors. 
 The spin current density can be as large as $2\times 10^8 Amps/cm^2$, and should be detectable in inelastic neutron scattering experiments with very cold neutrons\cite{exp}.
The excess charge near the surface as large as 1 electron per $10^6$ ions and the internal electric field as 
 large as $300,000 V/cm$ should be experimentally detectable.  It should be pointed out that the calculation of  electric fields in the neighborhood of non-spherical samples
 discussed earlier\cite{ellipsoid} needs to be modified in light of the results in this paper; this will be discussed elsewhere. Experimental verification of these predictions  will support the basic tenets of the theory of
 hole superconductivity: that superconductivity originates in the fundamental charge asymmetry of condensed matter,
 that it occurs when a metal has ``too many electrons'' (an almost full band, small de Broglie wavelength for the carriers at the Fermi energy, and with higher
 $T_c$ for anions\cite{narlikar}), and  that
 it is driven by kinetic energy lowering and involves dressed holes (antibonding electrons) turning into undressed electrons.
 Further development of the microscopic theory\cite{prb89,hsc} will be discussed in forthcoming work.

\acknowledgements
A helpful discussion with  B. Grinstein is gratefully acknowledged.

\end{document}